# A New Cluster-based Wormhole Intrusion Detection Algorithm for Mobile Ad-hoc Networks

[1]Debdutta Barman Roy, [2]Rituparna Chaki, [3]Nabendu Chaki

[1]Calcutta Institute of Engineering and Management, Kolkata, India,
barmanroy.debdutta@gmail.com
[2]West Bengal University of Technology, Kolkata 700064, India,
rchaki@ieee.org
[3]University of Calcutta, 92 A.P.C. Road, Kolkata 700009, India
nabendu@ieee.org

## ABSTRACT

*In multi-hop wireless systems, the need for cooperation among nodes to relay each other's packets exposes them to a wide range of security attacks. A particularly devastating attack is the wormhole attack, where a malicious node records control traffic at one location and tunnels it to another compromised node, possibly far away, which replays it locally. Routing security in ad hoc networks is often equated with strong and feasible node authentication and lightweight cryptography. Unfortunately, the wormhole attack can hardly be defeated by crypto graphical measures, as wormhole attackers do not create separate packets. They simply replay packets already existing on the network, which pass the cryptographic checks. Existing works on wormhole detection have often focused on detection using specialized hardware, such as directional antennas, etc. In this paper, we present a cluster based counter-measure for the wormhole attack, that alleviates these drawbacks and efficiently mitigates the wormhole attack in MANET. Simulation results on MATLab exhibit the effectiveness of the proposed algorithm in detecting wormhole attacks.*

## 1. INTRODUCTION

Mobile wireless ad hoc networks are fundamentally different from wired networks, as they use wireless medium to communicate, do not rely on fixed infrastructure, and can arrange them into a network quickly and efficiently. In a Mobile Ad Hoc Network (MANET), each node serves as a router for other nodes, which allows data to travel, utilizing multi-hop network paths, beyond the line of sight without relying on wired infrastructure. Security in such networks, however, is a great concern [1, 2, 7, 8]. The open nature of the wireless medium makes it easy for outsiders to listen to network traffic or interfere with it. Lack of centralized control authority makes deployment of traditional centralized security mechanisms difficult, if not impossible. Lack of clear network entry points also makes it difficult to implement perimeter-based defense mechanisms such as firewalls. Finally, in a MANET nodes might be battery-powered and might have very limited resources, which may make the use of heavy-weight security solutions undesirable [2, 3, 7, 8, 13].

A wormhole attack is a particularly severe attack on MANET routing where two attackers, connected by a high-speed off-channel link, are strategically placed at different ends of a network, as shown in figure 1. These attackers then record the wireless data they overhear, forward it to each other, and replay the packets at the other end of the network. Replaying valid network messages at improper places, wormhole attackers can make far apart nodes believe they are immediate neighbors, and force all communications between affected nodes to go through them.

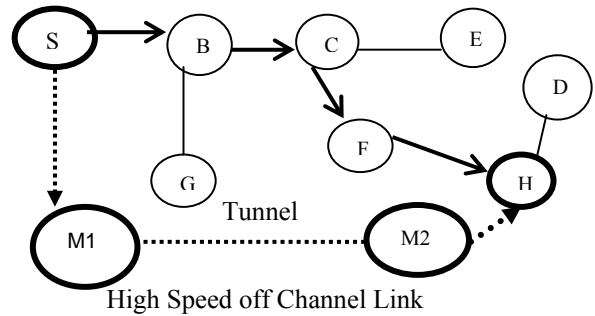

Figure 1: MANET with a wormhole attack

In general, ad hoc routing protocols fall into two categories: proactive routing protocol that relies on periodic transmission of routing packets updates, and on-demand routing protocols that search for routes only when necessary. A wormhole attack is equally worse a threat for both proactive and on-demand routing protocols [3, 7, 8, 9].

When a proactive routing protocol [10] is used, ad hoc network nodes send periodic HELLO messages to each other indicating their participation in the network. In Figure 2, when node S sends a HELLO message, intruder M1 forwards it to the other end of the network, and node H hears this HELLO message. Since H can hear a HELLO message from S, it assumes itself and node S to be direct neighbors. Thus, if H wants to forward anything to S, it may do so unknowingly through the wormhole link. This effectively allows the wormhole attackers full control of the communication link.

In case of on-demand routing protocols, such as AODV [11], when a node wants to communicate with another node, it floods its neighbors with requests, trying to determine a path to the destination. In figure 2, if S wants to communicate with H, it sends out a request. A wormhole, once again, forwards such request without change to the other end of the network, may be directly to node H. A request also travels along the network in a proper way, so H is lead to believe it has a possible route to node S thru the wormhole attacker nodes. If this route is selected by the route discovery protocol, once again wormhole attackers get full control of the traffic between S and H. Once the wormhole attackers have control of a link, attackers can drop the packets to be forwarded by their link. They can drop all packets, a random portion of packets, or specifically targeted packets1. Attackers can also forward packets out of order or 'switch' their link on and off [3].

In this paper, we have proposed an algorithm where intrusion detection has been done in a cluster based approach to detect the wormhole attacks. The AODV routing protocol is used as the underlying network topology. A two layer approach is used for detecting whether a node is acting as a wormhole.

## 2. RELATED WORKS

Routing security in ad hoc networks is often equated with strong and feasible node authentication and lightweight cryptography. A wide variety of secure extensions to existing routing protocols have been proposed over the years. However, the majority of these protocols are focused on using crypto graphical solutions to prevent unauthorized nodes from creating seemingly valid packets [8]. Unfortunately, the wormhole attack can not be defeated by crypto graphical measures, as wormhole attackers do not create separate packets - they simply replay packets already existing on the network, which pass all cryptographic checks. Perhaps the most commonly cited wormhole prevention mechanism is 'packet leashes' by Hu et al [13]. Hu proposed to add secure 'leash' containing timing and/or Global Positioning System (GPS) information to each packet on a hop-by-hop basis. Based on the information contained in a packet leash, a node receiving the packet would be able to determine whether the packet has traveled a distance larger than physically possible.

Hu proposed two different kinds of leashes: geographical leashes and temporal leashes. Geographic leashes require each node to have access to up-to-date GPS information, and rely on loose (in the order of ms) clock synchronization. When geographical leashes are used, a node sending a packet appends to it the time the packet is sent $t_s$ and its location $p_s$. A receiving node uses its own location $p_r$ and the time it receives a packet $t_r$ to determine the distance the packet could have traveled. Keeping in mind maximum possible node velocity v, clock synchronization error $\Delta$, and possible GPS distance error $\Delta$, the distance between the sender and the receiver $d_{sr}$ is upper-bounded by:

$$d_{sr} < \|p_s - p_r\| + 2v(t_r - t_s + \Delta) + \Delta$$

Geographical leashes should work fine when GPS coordinates are practical and available. However, modern GPS technology has significant limitations that should not be overlooked. While the price of GPS devices is going down, it remains substantial. Besides, GPS is somewhat of a nuisance for personal laptops. Also, while, as Hu [13, 3] specifies, it is possible to achieve GPS precision of about 3m with state-of-the-art GPS devices, consumer-level devices do not get (and do not require) this level of resolution. Finally, GPS systems are not versatile, as GPS devices do not function well inside buildings, under water, in the presence of strong magnetic radiation, etc. As opposed to geographical leashes, temporal leashes require much tighter clock synchronization (in the order of nanoseconds), but do not rely on GPS information. When temporal leashes are used, the sending node specifies the time it sends a packet $t_s$ in a packet leash, and the receiving node uses its own packet reception time $t_r$ for verification. In a slightly different version of temporal packet leashes, the sending node calculates an expiration time $t_e$ after which a packet should not be accepted, and puts that information in the leash. This is to prevent a packet from traveling farther than distance L

$t_e = t_s + L/C - \Delta$, where, C is the speed of light and $\Delta$ is the maximum clock synchronization error.

Another set of wormhole prevention techniques, somewhat similar to temporal packet leashes [10], is based on the time of flight of individual packets. Wormhole attacks are possible because an attacker can make two far-apart nodes see themselves as neighbors. One possible way to prevent wormholes, as used by Capkun et al [14], Hu et al [15], Hong et al [4], and Korkmaz [5], is to measure round-trip travel time of a message and its acknowledgement, estimate the distance between the nodes based on this travel time, and determine whether the calculated distance is within the maximum possible communication range. The basis of all these approaches is the following. The Round Trip Travel Time (RTT) $\delta$ of a message in a wireless medium can, theoretically, be related to the distance d between nodes, assuming that the wireless signal travels with a speed of light c:

$$d = (\delta c)/2 \text{ and } \delta = 2d/c$$

The neighbor status of nodes is verified if $d$ is within the radio transmission range $R$ for R > d (d within transmission range): $R > \delta c/2$ and $\delta < 2R/c$. In essence, the use of RTT eliminates the need for tight clock synchronization required in temporal leashes: a node only uses its own clock to measure time. However, this approach, while accounting for message propagation, completely ignores message processing time. When a message is sent by one node and is acknowledged by another, the time it takes for a node to process a message and to reply to it is generally non-negligible, particularly in the context of bounding short distances using signals whose speed is similar to that of light in vacuum. After all, it takes the light less than 0.2 seconds to circle the entire Earth around the equator. Outstanding clock precision and practically nonexistent errors are required to bind distances on the order of hundreds of meters.

Several researchers worked on the wormhole attack problem by treating a wormhole as a misbehaving link. In such approaches, a wormhole attack is not specifically identified. Rather, the wormhole's destructive behavior is mitigated. Baruch [6] and Chigan [12] use link rating schemes to prevent blackhole and wormhole attacks. They both rely on authenticated acknowledgements of data packets to rate links: if a link is dropping packets, the acknowledgements do not get through; link is rated low and avoided in the future. These approaches are geared towards discovery and prevention of only one kind of wormhole behavior: packet loss. Wormholes can do much more than that: they can send packets out of order, confuse location-based schemes, or simply aggregate packets for traffic analysis. Even the distortion of topology information that a wormhole introduce can be a significant problem in particular networks. The real problem with the wormholes is that unauthorized nodes (wormhole attackers) are able to transmit valid network messages. Techniques based on links' performance may be suitable in certain cases, but they do not fully address the wormhole problem.

In one of our earlier works, [1] a new collaborative algorithm called IDSX had been proposed. The proposed IDSX offers an extended architecture and is compatible with heterogeneous IDS already deployed in the participating nodes. In the high level of the architecture of the IDSX mechanism, the cluster heads act as the links across different clusters. The cluster heads are IDSX enabled and hence, can utilize alerts to generate the alarms. Alerts represent the potential security breaches as identified by local IDS active nodes. The IDSX nodes are authorized to take the final decision of discarding a node after aggregating and correlating the alerts that has been generated over a period of time.

## 3. PROPOSED METHODOLOGY

Our objective is to find out the malicious node that performs the wormhole attack in network. We have assumed that the MANET consists of clusters of nodes. The assumptions regarding the organization of the MANET are listed in section 3.1.

### 3.1 Assumptions

The following assumptions are taken in order to design the proposed algorithm.
1. A node interacts with its 1-hop neighbors directly and with other nodes via intermediate nodes using multi-hop packet forwarding.
2. Every node has a unique id in the network, which is assigned to a new node collaboratively by existing nodes.
3. The entire network is geographically divided into a few disjoint or overlapping clusters
4. The network is considered to be layered.
5. A cluster head at the inner layer is represented as CH (1,i), where 1 signifies inner Layer, and i stands for the cluster number
6. Each cluster is monitored by only one cluster head (monitoring node).
7. The cluster membership is restricted up to 2 hops.

### 3.2 Cluster formation

In this paper, we have proposed an algorithm where intrusion detection has been done in a cluster based manner to take care of the wormhole attacks. The AODV routing protocol is used as the underlying network topology. A two layer approach is used for detecting whether a node is participating in a wormhole attack. The layered approach is introduced to reduce the load of processing on each cluster heads. From security point of view, this will also reduce the risk of a cluster head being compromised.

The entire network is divided in clusters as in figure 2. The clusters may be overlapped or disjoint. Each cluster has its own cluster head and a number of nodes designated as member nodes. Member nodes pass on the information only to the cluster head. The cluster-head is responsible for passing on the aggregate information to all its members. The cluster head is elected dynamically and maintains the routing information.

GN is the guard node, used for monitoring the malicious activity. The main purpose of the guard node is to guard the cluster from possible attacks. The guard node has the power to monitor the activity of any node within the cluster. The guard node reports to the cluster head of the respective layer in case a malicious activity is detected. A cluster head in the inner layer (CH1,i) detects a malicious activity and informs the cluster head $CH_2$ of the outer layer to take appropriate action. It's the duty of (CH1,i) to check the number of false routes generated by any node. The cluster head $CH_2$ of outer layer takes upon itself the responsibility of informing all nodes of the inner layer about the malicious node.

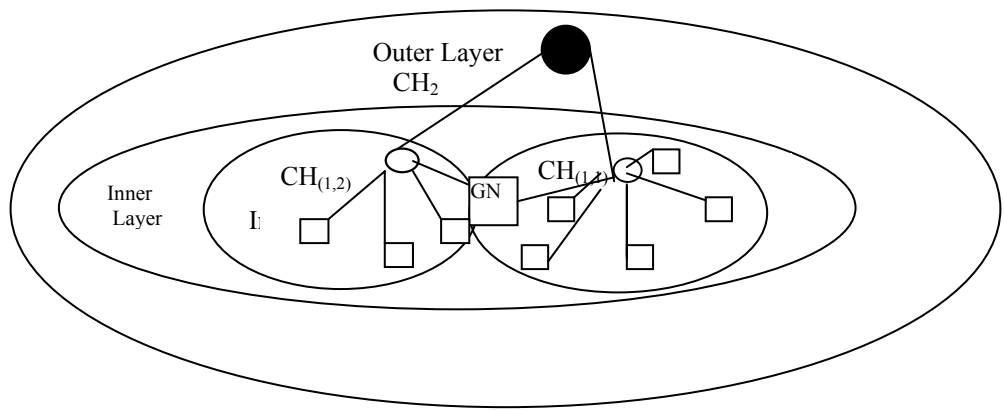

Figure 2 - The Layered structure

## 3.2 Cluster Based Detection Technique of Wormhole Attack in MANET

Before, we present the actual algorithm for detection of wormhole attacks, the data structure used for the purpose has been described below.
1. Round trip time ($T_r$): When the source node send packet it starts a timer. On receipt of an acknowledgement, the timer is stopped. The total time elapsed is recorded as Tr.
2. Expected time of delivery ($T_e$): The expected time of delivery of a packet to a destination node is calculated as the time taken when the source node send HELLO packet to the destination node and get back an acknowledgement for that.
3. Threshold tolerance ($P_{th}$): This refers to the threshold value defined by the monitoring node. It is the tolerance value for lost packets.
4. Neighbor table (Neighbor$_i$):Neighbor table for $i^{th}$ node consists of {neighbor_id} for all its neighbors.
5. PKTSNT (S, D): Number of packets sent to a destination node D from source node S.
6. PKTRCD (S, D): Number of packets received by node D from a specific source node S.

Next, we present the algorithm to detect wormhole attacks. When a node in the $i^{th}$ cluster of layer 1 suspect wormhole attack within the cluster, it informs the cluster head of $i^{th}$ cluster at layer 1, which is denoted as CH (1,i). CH(1,i) informs cluster head at layer 2 ($CH_2$), about the malicious node. $CH_2$ broadcast this information to all cluster heads at layer 1. The cluster heads at layer 1 inform their respective cluster members.

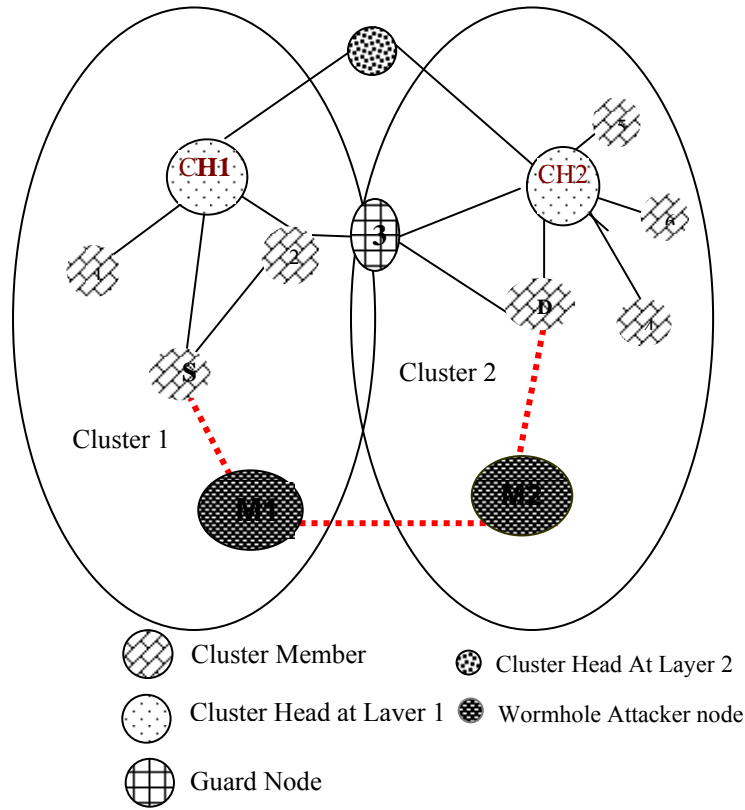

Figure 3: Cluster Based Detection Technique

In figure 3, node S sends a HELLO packet for destination node D. S has a path to D via (2, 3). M1, being in the proximity of S, overhears the HELLO message and forwards the same to node M2 in the other end of the network. Node D hears this HELLO message from S and therefore considers S to be its immediate neighbor and follow the route to send message to S via M1 and M2. The node 3 which is at the overlapping position of two cluster acts as GUARD node who can here every packet send by node S for the destination node D and monitor the packets route from souce to destination. The guard node is also called monitoring node. When S observes some malicious behavior when it sends packet to D it informs the guard node. The guard node then checks the number of packets send for the node D and those actually received by D from S. Then it calculates Δ*p* = *PKTSNT(S, D) - PKTRCD (S, D)*. If the value of Δ*p* surmounts the threshold value that is predefined by the monitoring node then monitoring node finds out the wormhole attack.

**Procedure WormHoleDetection**
Begin
    Step 1: Initiate the network with two cluster and each cluster have some nodes.
    Step 2: The node within a cluster having minimum node ID becomes Cluster Head. The node ID for each node is provided when the node enter into the cluster.
    Step 4: Each node stores the information of its immediate neighbors in its neighbor table.
    Step 5: The node nearest to both the cluster heads at layer 1 is chosen as the guard node.

Step 7: Source node S sends a HELLO packet to the intermediate node with destination node ID and cluster ID
    Step 7.1: S starts timer, initializes $T_1$
    Step 7.2: S increments the PKTCNT(S, D)
    Step 7.3: When S get acknowledgement from destination node stop timer, $T_2$
    Step 7.4: The expected round trip time is computed as $T_e = T_2 - T_1$
    Step 7.5: Source node S sends a packet to destination node
    Step 7.6: S starts timer $TP_1$
    Step 7.7: When S get acknowledgement from destination node stop timer, $TP_2$
    Step 7.8: The round trip time is calculated as $T_r = . TP_2 - TP_1$
    Step 7.9: If $T_r \ll T_e$ then inform guard node.
Step 8: The guard nodes checks number of packet send by source node PKTSNT (S, D) and number of packet receive by destination node PKTRCD(S, D).
Step 9: $\Delta p$ = PKTSNT (S, D) - PKTRCD (S, D).
Step 10: If $\Delta p > P_{th}$ then inform the source node to stop packet transfer.
Step 11: The source node stop packet transfer and inform cluster head.
End.

## 4. PERFORMANCE ANALYSIS

A simulation study has been done in MatLab. We have worked with a 30 nodes network while the number of packets sent is varied from 1 to 10. The number of guard nodes has been increased from 1 to 4. The following are the performance graphs of the network in the presence of 1, 2, 3, and 4 guard nodes.

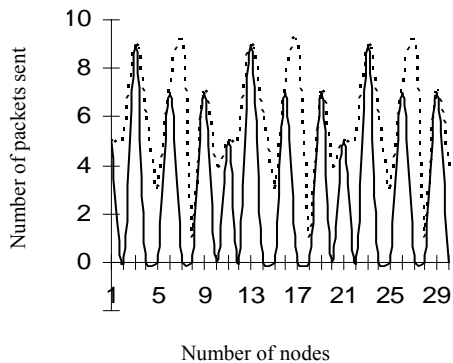

Figure 4: Number of packets sent and dropped in presence of 1 guard node

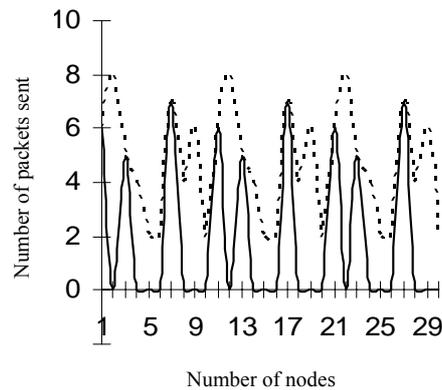

Figure 5: Number of packet sent and dropped in presence of 2 guard nodes

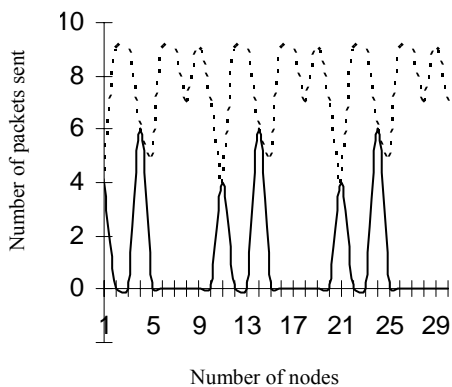

Figure 6: Number of packets sent and dropped in presence of 3 guard nodes

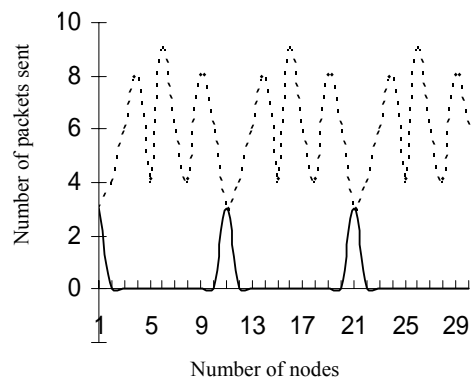

Figure 7: Number of packets sent and dropped in presence of 4 guard nodes

In figure 4, we observe that in presence of a single guard node, the number of packet drop and the number of packet send are nearly the same. So, there is a 50 % improvement in performance, with the presence of a single guard node.

In figure 5, the number of packet send and the number of packet drops vary due to the presence of 2 guard nodes. The performance of the network improves further. As evident from figures 6 and 7, it is observed that as the number of guard node increases to 3 and 4 respectively, the number of packets dropped minimizes. The performance of the network increases accordingly. Thus, the increase in the number of guard nodes steadily increases the probability of detection of wormhole attack.

## 5. CONCLUSION

In this work, a new cluster based wormhole detection method has been proposed. In multi-hop wireless systems, the need for cooperation among nodes to relay each other's packets exposes them to a wide range of security threats including the wormhole attack. A number of recent works have been studied before proposing this new methodology. The proposed solution unlike some of its predecessors does not require any specialized hardware like directional antennas, etc for detecting the attackers.  or extremely accurate clocks, etc. The simulation using 30 nodes and variable number of guard nodes prove the effectiveness of the proposed algorithm. Currently more studies are being done to analyze the performance of the proposed algorithm in presence of multiple attacker nodes.

**Authors**

Debdutta Pal received her M. Tech. Degree in Software Engineering from the West Bengal University of Technology in 2007. She is at present working as a Lecturer at Calcutta Institute of Engineering and Management, Kolkata, West Bengal. Her research interests include the field of Computer Networking, and Wireless Mobile Ad hoc Network.

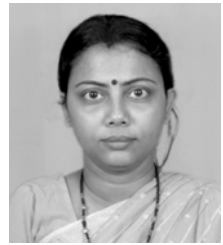

Rituparna Chaki is a Reader (Associate Professor) in the Department of Computer Science & Engineering, West Bengal University of Technology, Kolkata, India since 2005. She received her Ph.D. in 2002 from Jadavpur University, India. The primary area of research interest for Dr. Chaki is Wireless Mobile Ad hoc Networks. She has also served as a Systems Manager for Joint Plant Committee, Government of India for several years before she switched to Academia. Dr. Chaki also serves as a visiting faculty member in other leading Universities including Jadavpur University. Dr. Chaki has about 20 referred international publications to her credit.

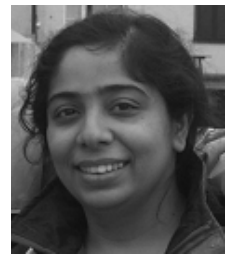

Nabendu Chaki is a faculty member in the Department of Computer Science & Engineering, University of Calcutta, Kolkata, India. He received his Ph.D. in 2000 from Jadavpur University, India. His areas of research interests include distributed computing and software engineering. Dr. Chaki has also served as a Research Faculty member in the Ph.D. program in Software Engineering in U.S. Naval Postgraduate School, Monterey, CA during 2001-2002. He is a visiting faculty member for many Universities including the University of Ca'Foscari, Venice, Italy. Dr. Chaki has published more than 50 referred research papers and a couple of text books. Besides being in the editorial board for 4 Journals, Dr. Chaki has also served in the committees of several international conferences.

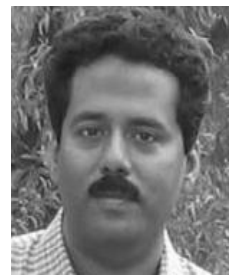